\documentstyle[epsf]{lya1}
\begin{document}

\load{\scriptsize}{\sc}
\def\ion#1#2{\rm #1\,\sc #2}
\def\HI{{\ion{H}{i}}}
\def\HII{{\ion{H}{ii}}}
\def\GI{{\ion{He}{i}}}
\def\GII{{\ion{He}{ii}}}
\def\GIII{{\ion{He}{iii}}}
\def\MH{{{\rm H}_2}}
\def\Hp{{{\rm H}_2^+}}
\def\Hm{{{\rm H}^-}}

\def\dim#1{\mbox{\,#1}}

\def\figname#1{lya1.#1}

\title[Hydrodynamics of the IGM]
{Probing the Universe with the Lyman-alpha Forest: \\I.
Hydrodynamics of the Low Density IGM}

\author[Gnedin and Hui]{Nickolay Y.\ Gnedin$^1$ and Lam Hui$^2$\\
$^1$Department of Astronomy, University of California, Berkeley, 
CA 94720\\
$^2$NASA/Fermilab Astrophysics Center, Fermi National Accelerator
Laboratory, Batavia, IL 60510}

\maketitle

\begin{abstract}

We introduce an efficient and accurate alternative to full hydrodynamic
simulations, Hydro-PM (HPM), for the study of the low
column density Lyman-alpha forest ($N_{\rm HI} \la 10^{14}\dim{cm}^{-2}$).
It consists of a Particle-Mesh (PM)
solver, modified to compute, in addition to the gravitational potential,
an effective potential due to the gas pressure. 
Such an effective potential can be computed from the density
field because of a tight correlation between density and pressure in the low
density limit ($\delta \la 10$), which can be calculated for any
photo-reionization history by a method outlined in Hui \& Gnedin (1997). 
Such a correlation exists, in part, 
because of minimal shock-heating in the low density limit. 
We compare carefully the density and velocity fields as well
as absorption spectra, computed using HPM versus hydrodynamic 
simulations, and find good agreement. 
We show that HPM is capable of reproducing  
measurable quantities, such as the column density distribution,
computed from full hydrodynamic simulations, to a precision
comparable to that of observations. 
We discuss how, by virtue of its speed and accuracy, HPM can enable us to use
the Lyman-alpha forest as a cosmological probe.

We also discuss in detail the smoothing of the gas (or baryon) fluctuation
relative to that of the dark matter on small scales due to
finite gas pressure: (1) It is shown the conventional
wisdom that the linear gas fluctuation is smoothed on the Jeans scale
is incorrect for general reionization (or reheating) history; the correct
linear filtering scale is in general smaller than the Jeans scale after
reheating, but larger prior to it. (2) It is demonstrated further that
in the mildly nonlinear regime, a PM solver, combined with suitable
pre-filtering of the initial conditions, can be used to model the
low density IGM. But such an approximation is shown to be less accurate
than HPM, unless a non-uniform pre-filtering scheme is implemented. 

\end{abstract}


\section{Introduction}

The low density intergalactic medium, filling the enormous space between 
galaxies and their aggregations, offers cosmologists a unique and powerful
probe of the high redshift universe ($z\sim2-5$), still inaccessible for 
large galaxy surveys. The intergalactic medium 
(hereafter IGM) manifests itself observationally in the numerous weak 
absorption lines along a line of sight to a distant quasar, 
the Lyman-alpha forest. 
Up to date, an enormous treasury
of observational data on the Lyman-alpha forest at a wide range of
redshifts has been collected (see, for example, 
Hu et al.\ 1995, Lu et al.\ 1996,
Cristiani et al.\ 1996, Kirkman \& Tytler 1997,
Kim et al.\ 1997 and D'Odorico et al.\ 1997 for most recent observational
advances). Several models were proposed
to explain the Lyman-alpha forest 
\cite{BS65,A72,B81,OI83,IO86,R86,I86,R88,BSS88,MG90,BBC92}. However, it was
only after several groups \cite{CMOR94,ZAN95,HKWM96,MCOR96,WB96,ZANM97}
performed cosmological hydrodynamic simulations
when it became apparent that at least an appreciable fraction of
the Lyman-alpha forest consists of smooth fluctuations in the IGM, which
arise naturally under gravitational instability,
rather than discrete absorbers, as has been believed before.

Abundance of observational data and the existence of a 
compelling theoretical framework (i.e. hierarchical clustering)
allows one to make detailed comparisons between observations
and predictions of a given cosmological model.
Moreover, one might even attempt to use the
observational data in a maximum-likehood type analysis to infer 
cosmological parameters, either in a model-independent way, or at
least within a class of models.
A recent example towards this direction is the use of
the observed mean Lyman-alpha optical depth to 
put limits on the baryon content of the universe 
\cite{MCOR96,RMS96,BD97,WMHK97}.
One might contemplate going a step further to use other properties of 
the Lyman-alpha forest as a probe of
equally interesting cosmological parameters such as 
the normalization and slope of the power spectrum (see, for example, 
Hui, Gnedin \& Zhang 1997), the
massive neutrino density, the epoch of reionization and so on.

However, while hydrodynamic simulations give us insights into
the physical properties of the IGM as well as definite predictions for a given
cosmological model (provided, numerical resolution and physical modelling 
are adequate), computational expense makes them impractical to use in
a maximum-likehood type of analysis in which a large range of
models are considered.

It is therefore important to ask whether a more efficient, and
at the same time sufficiently accurate, approximate method
can be developed in place of full hydrodynamic simulations.

Up to date, two different semi-analytical approximations have 
been used: the lognormal approximation \cite{BBC92,BD97,GH96}
and the (truncated) 
Zel'dovich approximation \cite{DS77,MG90,HGZ97}. While both approximations
are very efficient, they might not be sufficiently accurate. For
example, while the one-point density distribution function is close to 
lognormal for mildly nonlinear fluctuations, the lognormal approximation
itself fails to reproduce the density field accurately \cite{CMS93}.
The truncated Zel'dovich approximation is somewhat more accurate, and can be
used for  
about a decade in the IGM density, from about $\bar\rho/3$ to $3\bar\rho$, 
where $\bar\rho$ is the average density of the universe.
However, a main drawback of the truncated Zel'dovich approximation is
the necessity of the smoothing of the initial density field to 
minimize the amount of orbit-crossing by the time of interest.
This inevitably introduces artificial smoothing of small scale structure,
which could bias one's predictions, depending on the quantities one
is interested in.
While the Zel'dovich approximation can be successfully
applied to study the column density distribution of the Lyman-alpha forest
\cite{HGZ97}, it remains to be seen whether it can reproduce the detailed
results of a hydrodynamic simulation with sufficient accuracy.

In this paper we present a new approximate method, which is based on
a standard Particle-Mesh (PM) solver, modified to account for the pressure
forces acting on a fluid element. While the PM solver is significantly slower
than, say, the Zel'dovich approximation, it is still much faster than a
full hydrodynamic simulation. In order to develop a method that will be 
accurate to within 15\% (the reason for this number will be clear by
\S 4), we first in \S 2 describe two hydrodynamic simulations
that we have run to be used as templates against which approximate methods
are compared. Then, we start with linear theory to develop
some intuition first. In \S 3 we discuss the effect of the gas pressure
on the evolution of linear perturbations. The conventional wisdom
that linear baryon (or gas) perturbations are smoothed on the Jeans scale is 
shown to
be incorrect in general, and it is demonstrated 
that the smoothing scale depends on the reionization history of the universe.
Armed with an understanding of the behavior of linear fluctuations, 
in \S 4 we compare full hydrodynamic
simulations with an approximate method based on combining a PM solver
with the appropriate smoothing of initial conditions (with the smoothing scale
motivated by the linear analysis), as a way of taking into account the physical
effect of gas pressure (this is different from initial smoothing in the case
of the truncated Zel'dovich approximation as a way of correcting for 
orbit-crossing). 
We then conclude that this method is not sufficiently accurate
and proceed to develop our new approximation, which we call
Hydro-PM (hereafter HPM) in \S 5. 

The idea of HPM is very simple: one modifies a regular PM solver
to compute, in addition to the usual gravitational potential,
an effective potential due to the presence of gas pressure. 
This is possible because there exists a tight correlation between
temperature and density (or equivalently, between pressure and density)
in the low density limit, which can be computed quite accurately 
for any given reionization history \cite{HG97}.
A given 
density field then predicts an effective pressure field as well as a
definite gravitational potential field.
The fundamental rationale is that for the Lyman-alpha forest of
sufficiently low column density ($N_{\rm HI}\la10^{14}\dim{cm}^{-2}$), 
shock-heating is not important 
(or, equivalently, the density fluctuations
are only mildly nonlinear, $\delta \la 10$), which is
the one piece of physics that HPM does not incorporate.
As we will show, this does not compromise our accuracy significantly
while buying us a great increase in efficiency over full
hydrodynamic simulations.

It is appropriate that we mention here two wonderful pieces of related work. 
Petitjean, M\"{u}cket \& Kates \shortcite{PMK95} and M\"{u}cket et 
al.\ \shortcite{MPKR96}
investigated properties of the Lyman-alpha forest using PM simulations,
suitably modified to follow the thermodynamics of baryons. 
Their approach differ from our HPM method in at least two aspects:
the baryons are approximated as following trajectories of dark
matter prior to shell-crossing (we include the dynamical effects
of pressure on baryons), and shock-heating is modelled in their
method which enables them to study higher column density systems. 
The reader is referred to the above papers for details.

Finally, we conclude in \S 6 with a brief
discussion. A word on our notation: the symbol $\rho$ is used to denote
mass density as usual, as well as the mass density in units of the
cosmic mean (i.e. $\rho$ and $1+\delta$ used interchangeably). 
Which meaning is intended should be clear from the context.

\section{Hydrodynamic Simulations}

\begin{table}
\label{tabmod}
\caption{Cosmological Models}
\medskip
\[
\begin{tabular}{ccccccc}
Model & $\Omega_0$ & $\Omega_b$ & $h$ & $\Omega_\Lambda$ & $\sigma_8$ & 
cell size\\
\noalign{\hrule}
LCDM &  0.35       & 0.055      & 0.7 & 0.65 & 0.79 & $6.6h^{-1}\dim{kpc}$ \\
SCDM &  1.0        & 0.05       & 0.5 & 0    & 0.70 & $15.6h^{-1}\dim{kpc}$\\
\end{tabular}
\]
\end{table}
We have performed two cosmological hydrodynamic simulations
against which we will measure the performance of our approximate
methods.
We used the SLH-P$^3$M code as described by Gnedin (1995), Gnedin (1996),
Gnedin \& Bertschinger (1996) and Gnedin \& Ostriker (1997).
Table \ref{tabmod} contains cosmological parameters
of the two models. We have chosen two different models which enable us
to test our approximate methods under different conditions.
For both models we have used $64^3$ baryonic mesh and the softening 
parameter was set to $1/3$, which gives us a dynamical range of $192$. Since we
are mainly concerned with modelling the low density IGM, $\delta\la10$,
we do not need to set the softening parameter to a very small value.
Our choice of the softening parameter, however, does enable
full resolution of regions with $\delta=10$ or lower.
The LCDM model is identical to
the model used in our Equation of State paper \cite{HG97}, except
for the larger value of the softening parameter, while the SCDM model
is close to the model studied by Zhang et al.\ \shortcite{ZANM97}.

\def\capJZ{
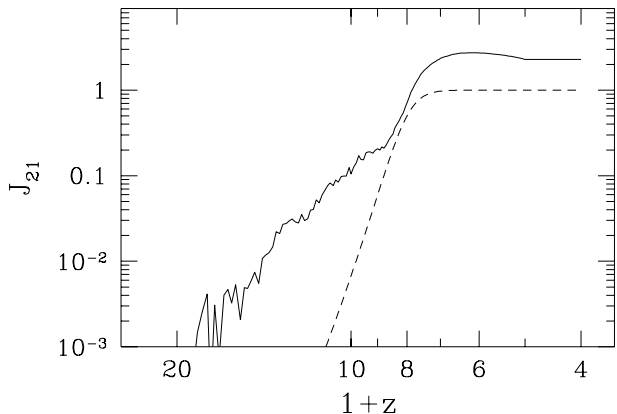
Evolution of the ionizing intensity $J_{21}$ for the LCDM model
({\it solid line\/}) and the SCDM model ({\it 
dashed line\/}) as a function of 
redshift.
}
\begin{figure}
\par\centerline{%
\epsfxsize=1.0\columnwidth\epsfbox{\figname{figJZ.ps}}}%
\caption{\label{figJZ}\capJZ}
\end{figure}
The thermal history for each of our simulations is determined by
the evolution of the photo-ionizing background. Figure \ref{figJZ} shows
the evolution of the ionizing intensity $J_{21}$ as a function of redshift
for both hydrodynamic simulations. The redshift evolution of $J_{21}$  
and the spectrum of radiation for the LCDM model was adopted from the 
simulation described in \cite{HG97}. For the SCDM model, we have adopted the 
following form of the redshift evolution of $J_{21}$:
\[
	J_{21}(z) = {1\over2}
	\left[1+\tanh\left(100{7-z\over8(1+z)}\right)
	\right],
\]
where $J_{21}$ is defined in exactly the same way as $J_{\rm HI}$ in 
Hui \& Gnedin \shortcite{HG97}, and we adopt the same spectral shape
as in Hui \& Gnedin \shortcite{HG97}, equation (7).
This form of the redshift evolution of $J_{21}$ is close to the sudden 
reionization models discussed in detail in Hui \& Gnedin (1997). For low $z$, 
$z\ll7$, the ionizing intensity reaches its asymptotic value, $J_{21}=1.0$, 
and it drops quickly before the redshift of reionization, $z_{\rm rei}=7$. 
The factor of 100 inside the $\tanh$  insures that reionization occurs 
smoothly in a redshift interval $\Delta z/z\sim0.01$.
This transition period is introduced to avoid numerical instabilities 
possible when $J_{21}$ increases abruptly at the redshift of reionization, 
as in models of sudden reionization. 

Both simulations have been continued until $z=3$. It is worth pointing out 
that the 
simulation box in both cases was rather small, $1h^{-1}\dim{Mpc}$ for the 
LCDM model, and $422h^{-1}\dim{kpc}$ for the SCDM model. At the final 
redshift,
fluctuations at the box size are already nonlinear, and simulation boxes are 
not representative patches of the universe. This fact should have
minimal effect on the present 
work: our main goal is to develop an understanding of the relationship between
the dark matter and the baryons on small scales, to help us find
an approximation that takes into account both gravity and gas pressure in
an appropriate manner (on large scales, things are simple: dark matter
and baryons simply trace each other). 
The key is then to resolve structure on the relevant
small (mass) scales (as will be explained in the next section),
rather than having a representative sample of the universe on large
scales.

\section{Linear Evolution of Cosmological Perturbations}

The linear evolution of perturbations in the dark matter - baryon
fluid is governed by two second order differential equations:
\[
	{d^2\delta_X\over dt^2} + 2H{d\delta_X\over dt} = 
	4\pi G\bar\rho(f_X\delta_X+f_b\delta_b),
\]
\begin{equation}
	{d^2\delta_b\over dt^2} + 2H{d\delta_b\over dt} = 
	4\pi G\bar\rho(f_X\delta_X+f_b\delta_b) - 
	{c_S^2\over a^2} k^2 \delta_b,
	\label{let}
\end{equation}
where $\delta_X(t,k)$ and $\delta_b(t,k)$ are Fourier components of 
density fluctuations 
in the dark matter 
and baryons (we equate baryons with the cosmic gas in this paper) 
respectively, 
which have respective mass fractions $f_X$ and 
$f_B$, $H(t)$ is the Hubble constant, $a(t)$ is the cosmological
scale factor, $\bar\rho(t)$ is the
average mass density of the universe, $c_S(t)$ is the sound speed in the cosmic
gas (where the sound speed is simply defined by $c_S^2 \equiv dP/d\rho$, 
assuming an equation of state that relates the $P$ and $\rho$), $k$ is the
comoving wavenumber and $t$ is the proper time.  

In the limit where the dark matter is gravitationally dominant, $f_b=0$ in
equation (\ref{let}), the growth of $\delta_X$ is described by the 
familiar factor $D_+(t)$ \cite{P80}, 
which coincides with $a(t)$ if the matter density
of the universe is critical.

The right hand side of the equation for $\delta_b$ contains two terms: the
gravitational force and the pressure force. On large scales, in the
limit $k\rightarrow0$, the pressure force can be neglected, and the baryon
fluctuation obeys the same equation as the dark matter fluctuation.
Assuming that $\delta_b = \delta_X$ initially, we have
\[
	\delta_b(t,k\rightarrow0) = \delta_X(t,k\rightarrow0) \propto D_+(t).
\]
On small scales, $k\rightarrow\infty$, the pressure force is dominant,
and one would expect that the baryon fluctuation is suppressed compared
to the dark matter fluctuation. A characteristic scale, where the two
forces are equal, is called the Jeans scale. We denote the wavenumber
corresponding to the Jeans scale as $k_J$,
\begin{equation}
	k_J = {a \over c_S}\sqrt{4\pi G\bar\rho}.
	\label{defkj}
\end{equation}
The Jeans scale is in general a function of time, but for the special
case when the gas temperature $T$ evolves with time as $T \propto 1/a$,
the Jeans scale is constant in time. In this case, and assuming $f_b=0$
(i.e.\ the baryons are gravitationally subdominant),
so that $\delta_X \propto D+$, equation (\ref{let}) can be solved analytically:
\begin{equation}
	\delta_b(t,k) = {\delta_X(t,k)\over 1 + k^2/k_J^2}.
	\label{tacase}
\end{equation}
Thus, at small scales, $k\rightarrow\infty$, the baryon fluctuation is
suppressed relative to the dark matter fluctuation by a factor of 
$k_J^2/k^2$.
Note that this assumes effectively a very special boundary
condition: $T\propto1/a$ {\it at all times}.

Let us now consider a more realistic case. At sufficiently high redshifts,
before the Compton heating of the baryon gas by the CMB becomes
inefficient, the evolution of the baryon temperature is well described
by the $T\propto1/a$ law. At redshifts of about $100$ (depending on the baryon
density of the universe), 
Compton scattering becomes inefficient and the temperature of the
baryon gas drops adiabatically, $T\propto1/a^2$. 
By low redshifts, but before the universe reionizes, the gas
temperature can be treated as effectively zero (or in other words,
the Jeans mass associated with the CMB temperature is too small
to be relevant for the modelling of the Lyman-alpha forest). 

The gas temperature rises dramatically after the universe
reionizes, and its subsequent evolution is obviously our object of interest.
As we will show, equations (\ref{defkj}) and (\ref{tacase})
no longer provide a correct description of the smoothing
and time evolution of the gas.

In order to consider a realistic case, we extract the evolution 
of the sound speed from our SCDM simulation and solve equation (\ref{let}) 
numerically with the obtained form of $c_s(t)$.

\def\capDB{
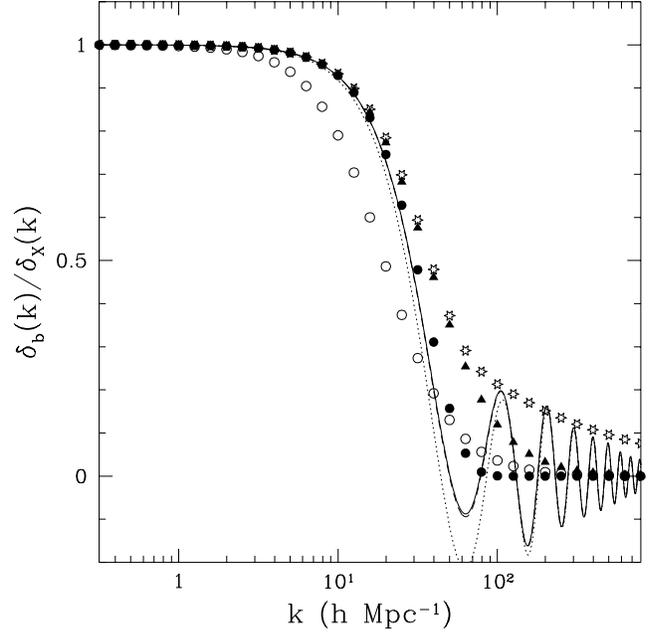
Comparisons of different filtering.
The exact linear baryon fluctuation for the SCDM model at $z=3$ 
as calculated from
equation (\protect{\ref{let}}) are shown for $f_b=0.05$
({\it solid line\/}), $f_b=0$ ({\it dashed line\/}, almost overlapping with the
{\it solid line\/}), and $f_b=1$ 
({\it dotted line\/}). Points with different symbols
represent different filtering of the linear dark matter fluctuation
(to approximate the linear baryon fluctuation):
$1/(1+k^2/k_J^2)$ filtering ({\it open circles\/}),
$\exp(-k^2/k_F^2)$ filtering ({\it filled circles\/}),
$1/(1+k^2/k_F^2)$ filtering ({\it filled triangles\/}), 
and a hybrid filtering which gives the best fit to the envelope of
the baryon fluctuation (eq.\ 
[\protect{\ref{bestfitfil}}]; {\it stars\/}).
}
\begin{figure}
\par\centerline{%
\epsfxsize=1.0\columnwidth\epsfbox{\figname{figDB.ps}}}%
\caption{\label{figDB}\capDB}
\end{figure}
Figure \ref{figDB} shows the linear baryon fluctuation, normalized by
the linear dark matter fluctuation, at $z=3$.
Three different cases are shown: $f_b=0.05$ as in the simulation 
(the solid line), $f_b=0$ in equation
(\ref{let}), baryons being treated as gravitationally
subdominant and therefore their gravitational effect is negligible 
(the dashed line), and 
$f_b=1$, all matter being baryonic  
(the dotted line). The two cases with $f_b=0.05$ and $f_b=0$ can 
barely be distinguished from each other in the figure.
The case where all the matter is treated as baryonic also gives a very similar 
result. This point will be examined further in \S 6. 

Let us focus on the open 
circles for the time being. They represent the linear baryon
fluctuation as given by equation (\ref{tacase}), where the
filtering scale is the Jeans scale as defined in equation (\ref{defkj})
for $z=3$.
One can see that in this realistic case (where $T$ does not evolve
as $1/a$ at all times), filtering of the baryon fluctuation 
occurs at a smaller scale than the Jeans scale, contrary to conventional
wisdom. In addition, oscillations occur at small scales,
and the amplitude of these oscillations decay at a rate slower that $1/k^2$,
contrary to equation (\ref{tacase}).

It is possible to understand this analytically. Let us 
consider the case where the baryons are gravitationally subdominant, $f_b=0$. 
Then the dark matter fluctuation simply grows like
$D_+(t)$ (ignoring the decaying mode). Let us consider expanding the 
ratio of the baryon fluctuation to dark matter fluctuation
$\delta_b(t,k)/\delta_X(t,k)$ in powers of $k^2$. Retaining only the first 
two dominant terms in the small $k$ limit, and recalling that
$\delta_b(t,k=0)=\delta_X(t,k=0)$, we 
have: 
\begin{equation}
	{\delta_b(t,k)\over \delta_X(t,k)} = 1 - {A(t)\over D_+ (t)}k^2,
	\label{delbtwok}
\end{equation}
where $A(t)$ is an unknown coefficient to be determined. Inserting equation 
(\ref{delbtwok}) into (\ref{let}) and ignoring terms of order $k^4$ or higher, 
we obtain the following equation for $A(t)$:
\begin{equation}
	{d^2A\over dt^2} + 2H{dA\over dt} =  {c_S^2\over a^2}D_+(t).
	\label{eqA} 
\end{equation}
This equation can be easily solved by:
\begin{equation}
	A(t) = \int_0^t dt^\prime c_S^2(t^\prime) D_+(t^\prime) 
	\int_{t^\prime}^t
	{dt^{\prime\prime}\over a^2(t^{\prime\prime})}.
\end{equation}
where the initial conditions $A(t=0) = dA/dt(t=0) = 0$ are assumed
(i.e. no difference between the baryon and dark matter fluctuations
at early times).
Note that $A$ is positive, which means the baryon fluctuation is {\it always\/}
suppressed, compared to the dark matter, in the low $k$ regime.
We now introduce the {\it filtering scale\/}, with the corresponding
wavenumber denoted as $k_F$, by the following 
expression:
\[
	A(t) \equiv {D_+(t)\over k_F^2(t)},
\]
so that equation (\ref{delbtwok}) can now be rewritten as
\begin{equation}
	{\delta_b(t,k)\over\delta_X(t,k)} = 1-{k^2\over
	k_F^2}. 
	\label{defkf}
\end{equation}
The following expression for the filtering scale $k_F$ can be obtained:
\begin{eqnarray}
	{1\over k_F^2(t)} & = & {1\over D_+(t)} \int_0^t dt^\prime 
	a^2(t^\prime)
	{\ddot{D}_+(t^\prime)+2H(t^\prime)\dot{D}_+(t^\prime) 
	\over k_J^2(t^\prime)} \nonumber \\
	& & \int_{t^\prime}^t{dt^{\prime\prime}\over a^2(t^{\prime\prime})},
	\label{kfaskj}
\end{eqnarray}
where we have replaced the sound speed by its expression in terms of the Jeans
scale 
at the same moment (eq.\ [\ref{defkj}]), and dot denotes differentiation with
respect to the time $t$. 

An important conclusion follows from equation (\ref{kfaskj}).
Let us rewrite it in the following form, using the median value
theorem:
\begin{eqnarray}
	{1\over k_F^2(t)} & = &
	{1\over k_J^2(t_*)}\left[{1\over D_+(t)} \int_0^t 
	dt^\prime a^2(t^\prime)\right. \nonumber \\
	& & \left. 
	\left(\ddot{D}_+(t^\prime)+2H(t^\prime)\dot{D}_+(t^\prime)\right)
	\int_{t^\prime}^t	{dt^{\prime\prime}\over 
	a^2(t^{\prime\prime})}\right]\nonumber,
\end{eqnarray}
where $t_*$ is between 0 and $t$. The expression in square brackets 
integrates to 1, and we obtain:
\begin{equation}
	k_F(t) = k_J(t_*),
	\label{kfkj}
\end{equation}
where $t_*\leq t$. In other words, the filtering scale at a given time is
equal to the Jeans scale at
some earlier time. In particular, if the Jeans scale $1/k_J$ is an increasing
function 
of time, which is typically the case for sufficiently low redshifts after
reionization, the
filtering scale $1/k_F$ is always {\it smaller\/} than the Jeans scale.
The reverse would be true prior to reionization, as we will see in a moment.

The above notion of the filtering scale is, strictly speaking, only applicable
in the small $k$ limit, because it is derived based on an expansion in $k^2$ 
(equation [\ref{delbtwok}]). 
To see how well this filtering scale provides a description of
the linear baryon fluctuation in the high $k$ regime, we show in
Fig.\ \ref{figDB} with filled circles the filtering in the form
$\exp(-k^2/k_F^2)$ 
(i.e. $\delta_b = \delta_X \exp[-k^2/k_F^2]$), where $k_F$
is computed from equation (\ref{kfaskj}) using the evolution of the sound speed
(or in other words the
Jeans scale) as extracted from the SCDM hydrodynamic simulation. 

One can see
despite the fact that $k_F$ is derived in the small $k$ limit, the 
exponential filtering with $k_F$ gives an excellent fit to the baryon
fluctuation even for high $k$, until
oscillations take over. We also show with filled triangles the filtering of the
form $1/(1+k^2/k_F^2)$ (i.e. $\delta_b = \delta_X/[1+k^2/k_F^2]$) for 
the same $k_F$, which
gives a worse fit for the high $k$ cut-off and, as in the case of
$\exp(-k^2/k_F^2)$ filtering, does not 
match the envelope of oscillations on small scales.

Encouraged by the excellent performance of the gaussian filtering on
scale of $1/k_F$ in reproducing the 
exact linear solution, we now consider a few special cases, where 
$k_F$ can be calculated analytically. Let us
restrict ourselves  
to an $\Omega_0=1$ universe, where $D_+(t)=a(t)$. For simplicity, we will 
assume that
the mean molecular weight of the cosmic gas does not change, in which case
the sound speed is directly proportional to the square root of the gas
temperature. First, we consider the case where the gas temperature $T$ is zero
before reionization (which occurs at $a=a_{\rm rei}$), and remains constant
thereafter:
\begin{equation}
	T = \cases{0,& $a<a_{\rm rei}$, and \cr
		T_0,& otherwise.}
	\label{casea}
\end{equation}
Computing the integral (\ref{kfaskj}), we obtain for $a > a_{\rm rei}$:
\begin{equation}
	{1\over k_F^2} = {1\over k_J^2}{3\over10}\left[1 +
	4\left(a_{\rm rei}\over a\right)^{5/2} - 
	5\left(a_{\rm rei}\over a\right)^2\right].
	\label{kftconst}
\end{equation}
In particular, for $a\gg a_{\rm rei}$, 
\[
	k_F = \sqrt{10\over3}k_J.
\]

Another instructive example is when the gas temperature decays as $1/a$ after
reionization,
\begin{equation}
	T = \cases{0,& $a<a_{\rm rei}$, and \cr
		T_0 a_{\rm rei}/a,& otherwise.}
	\label{caseb}
\end{equation}
In this case the filtering scale for $a>a_{\rm rei}$ is given by
\begin{equation}
	{1\over k_F^2} = {1\over k_J^2}\left[1 +
	2\left(a_{\rm rei}\over a\right)^{3/2} -
	3{a_{\rm rei}\over a}\right].
	\label{kftaconst}
\end{equation}
In the limit $a\gg a_{\rm rei}$ we recover the standard result $k_F=k_J$,
but the asymptote is reached only slowly, and even at $z=3$ and for
$z_{\rm rei}=7$, we obtain $k_F = 2.2 k_J$.
We emphasize the departure of the correct filtering scale from the
usual Jeans scale is a result of the fact that $T$ above is not
assumed to evolve as $1/a$ at all times. The time evolution of $T$
considered above is partly motivated by reionization models
in which the originally cool cosmic gas was heated up to a high 
temperature by radiation emitted by sources (stars, quasars, etc) that turned
on at some high redshift.

Typically, the gas temperature decays as an intermediate power between
$a^0$ and $a^{-1}$ after reionization \cite{HG97}. We, therefore, conclude that
in a realistic  
case one should expect that at $z\sim3$ the filtering scale of the cosmic 
gas is about a factor of $1.5-2.5$ smaller than the Jeans scale, unless the 
universe reionized at a very high redshift, $z_{\rm rei}\gg10$.

Another interesting example is the evolution of the baryon perturbations
before reionization. After recombination at $z\sim1200$, the cosmic
gas temperature is still coupled to the CBR temperature by Compton heating,
and therefore evolves as $T\propto1/a$.
At a later time $a_{\rm dec}=0.01(\Omega_bh^2/0.0125)^{2/5}$, Compton heating 
becomes inefficient, and the gas temperature decreases adiabatically,
$T\propto1/a^2$. Since the Jeans scale decreases with time for an adiabatically
cooling gas, the filtering scale for the cosmic gas is actually
{\it larger\/} than the Jeans scale. More precisely, a good approximation
to the evolution of the cosmic gas temperature is given by the following
expression:
\begin{equation}
	T = \cases{2.73\dim{K}/a, & $a<a_{\rm dec}$, and \cr
		2.73\dim{K}a_{\rm dec}/a^2 & otherwise.}
	\label{casec}
\end{equation}
In this case the filtering scale for $a>a_{\rm dec}$ is given by
\begin{equation}
	{1\over k_F^2} = {1\over k_J^2}\left[3 \, {\rm ln}(a/a_{\rm dec})-3
	+ 4\left(a_{\rm dec}\over a\right)^{1/2}\right].
	\label{kfdec}
\end{equation}
For example, for $\Omega_bh^2=0.0125$, $k_F=0.45 k_J$ at $z=10$, and in
term of masses, the characteristic mass scale on which the gas distribution is
smoothed, $M_F\propto1/k_F^3$, is about 11 times {\it larger\/} than the Jeans
scale, $M_J\propto1/k_J^3$. This result has important implications for
understanding the formation of the first bound objects in the universe.

Next, we turn our attention to the oscillations in the high $k$ regime,
a behavior we can understand analytically for the time evolution 
specified in equation (\ref{caseb}). We can solve equation (\ref{let}) 
exactly in this case, assuming once again 
the case of an $\Omega_0=1$ universe with $f_b=0$ (i.e. baryons being
gravitationally subdominant): 
\[
	{\delta_b(t,k)\over \delta_X(t,k)} = {1\over1 + k^2/k_J^2}\left(1 + 
	{k^2\over k_J^2}\left[
	{n_-\over n_- -n_+}\left(a\over a_{\rm rei}\right)^{n_+} - \right.
	\right.
\]
\begin{equation}
	\phantom{AAAAA} \left.\left.
	{n_+\over n_- -n_+}\left(a\over a_{\rm rei}\right)^{n_-}\right]\right)
	\label{fullsol}
\end{equation}
for $a>a_{\rm rei}$, where
\[
	n_\pm = -{5\over 4} \pm {3\over 4}\sqrt{{1\over 9} - {8\over 3} 
	{k^2\over k_J^2}},
	\label{npm}
\]
and where $\delta_X$ grows as $a$. 
Note that since $T\propto1/a$ at $a>a_{\rm rei}$, the Jeans scale $k_J$ 
is constant in time. 
In the limit $a\gg a_{\rm rei}$ and for $k$ sufficiently small, equation
(\ref{fullsol}) reduces to equation  
(\ref{tacase}).

Let us now consider a fixed final $a$, and take the large $k$ limit. Then
both  $n_+$ and $n_-$ become complex (but $\delta_b$ is still real), and 
$\delta_b$ as a function of $k$ oscillates. However, one can see that in 
the high $k$ limit, the amplitude of these oscillations is independent of
$k$. We,  
therefore, conclude that in general $\delta_b /\delta_X$ has no
power-law asymptote in the high $k$ limit. It is sometimes 
claimed in the literature that $\delta_b /\delta_X$ always
approaches an asymptote of $k^{-2}$ in the high $k$ limit. 
That statement is only correct if $T$ evolves as a fixed power law in $a$
{\it at all times} (see Bi et al. 1992 for derivation).
The simple case above provides an example of departure
from this property.

Finally, we emphasize that the two hydrodynamic simulations described
in the previous section have sufficiently small cell sizes so that
the corresponding correct filtering scales ($1/k_F$) are resolved by
about 5 mesh cells. This ensures that we can meaningfully compare
different smoothing prescriptions, as explained in the following
section.

\section{Filtering Initial Conditions for a PM Simulation}

The linear analysis in the previous section shows that
the two mass components, the dark matter and the 
cosmic gas, evolve differently on small scales: the dark matter is affected by
gas pressure only via gravitational interaction with the 
gas, while the gas evolution is directly influenced by the thermal pressure 
on sufficiently small scales. In order to compute this complex 
interaction in every detail, a two-component hydrodynamic simulation is needed.
But often the precision achieved by the full hydrodynamic simulation is not 
required. For instance, current observations of the Lyman-alpha forest
typically give about 10-15\% accuracy for the column density distribution. We
will attempt to develop an approximation, that is significantly faster
than a hydrodynamic simulation, but at the same time gives us results with
similar accuracy.

As a step toward this goal, we will concentrate in this paper
on single-component approximations, i.e.\ approximations where the evolution 
of the cosmic gas is computed using only one set of resolution elements
(in our case particles) instead of following both the dark matter and the
gas separately. This approach is certainly more economical than a full
hydrodynamic simulation, but the question is: can we
make it accurate enough?

It is certainly possible to emulate a gas-dynamic solver using a simple
dark matter solver in the linear regime.
Let $\delta_X^{(0)}(t,k)$ and $\delta_b^{(0)}(t,k)$ be 
the linear solutions
to equation (\ref{let}) for a specific cosmological model. Suppose
we are interested in the baryon fluctuation at some final moment $t=t_f$
(which is early enough so that the fluctuation remains linear).
Let us model the evolution of the baryon perturbation with
a dark-matter-only solver (e.g. PM), which, in the linear regime, is 
equivalent to solving the first of equations
(\ref{let}) and assuming $f_b=0$. If
we choose the following initial condition for the dark-matter-only solver 
at an early time $t=t_i$: 
\begin{equation}
	\delta_X(t=t_i,k) = {D_+(t_i)\over D_+(t_f)}\delta_b^{(0)}(t=t_f,k),
	\label{xasblin}
\end{equation}
it is easy to see that we will reproduce the baryon
fluctuation in the linear regime at $t=t_f$. Since, as we have shown in the
last section, $\delta_b^{(0)}(t=t_f,k)$ 
can be modelled by $\delta_X(t=t_f,k)$ multiplied by some suitable filter, 
the above initial condition is equivalent to smoothing the initial
$\delta_X$ with the same filter.

One might then hope to model the dynamical evolution of 
the gas by a PM simulation, with the initial conditions
appropriately smoothed. In other words, one may try to model the gas 
evolution under the assumption that the gas is influenced by
gravity alone, hoping that the initial
filtering procedure is sufficient to model the
effect of pressure.

There is no guarantee that this simple-minded method would work.
After all, our idea of a simple filtering scale is derived
from linear analysis, while for our applications, we are
interested in regions of space with overdensity below, but
reaching up to about $10$. 
In fact, we will show in this section that this method
works to a certain extent, but is {\it not} good enough,
i.e.\ it fails to achieve an accuracy of
$10-15\%$ in a point-by-point
comparison of density and velocity fields against full hydrodynamic 
simulations. Observationally, interesting quantities such as
the column density distribution are typically measured with an 
accuracy of about $10-15\%$. As we will show in the next section, this level
of accuracy requires similar accuracies in the density and velocity fields
themselves. 

Before we embark upon a quantitative comparison of the PM + filtering 
method versus hydrodynamic simulation, we have to address
one technical point.

A collisionless (alias ``N-body'') numerical simulation, 
such as PM, uses particles to follow the evolution of the system.
For our applications, it is eventually necessary to compute the gas density
and velocity as a function of spatial positions.
How does one convert a distribution of particles into, say, the density field?
There exist several 
techniques, but in this paper we will adopt the simplest method of assigning
the density onto a uniform mesh using 
particle weights. Specifically, we will use the Triangular-Shape-Clouds
(TSC) scheme to assign the particle density onto a mesh. This method, however,
suffers from numerical noise. For example, in a sufficiently underdense region
a particle might be so remote from its neighbors that the TSC
assignment would leave empty regions (zero density) between the particle and
some of its neighbors. This generates unphysical structure on small scales.
The easiest way to suppress this numerically generated structure is to smooth
the resultant density distribution with, for example, a gaussian filter.
However, since we are trying to achieve an order of 10-15\% accuracy in
reproducing the gas distribution, 
we ought to ensure that we reduce the numerical
assignment noise to within a couple of percent. In other words, what is
the degree of smoothing we must apply to the TSC-assigned density distribution
to reduce the noise to, say, 2\%?

In order to answer this question, we have performed two PM simulations: a 
low resolution one with $64^3$ mesh, and a high resolution 
one with $128^3$ mesh with the same number of particles ($64^3$) and
identical initial conditions. 
The rms overdensity
at the final moment is chosen to be 3 to allow for development of sufficient
nonlinearities. 
Those two simulations
therefore should produce the same final density distribution except that the
high resolution simulation has twice higher spatial resolution. The two density
distributions are then smoothed with a gaussian filter with some
chosen smoothing scale.
By comparing the
two simulations smoothed with varying smoothing scale, 
we find that the smoothing 
scale should be at least 3 cells to reduce the numerical assignment noise
to within 2\%. This conclusion has also been confirmed by Bertschinger
\shortcite{B97}.

Therefore, from now on, we will present results of collisionless
N-body experiments with the final density and velocity fields assigned to
the uniform mesh by the TSC scheme
and then smoothed with a gaussian filter of three mesh cells.
This procedure is admittedly quite wasteful, since it implies that
we lose a factor of 1.5 to 3 in spatial resolution. The advantage
is that it is simple to implement. We defer developing
a more efficient density assignment scheme to future work.  

\def\capNH{
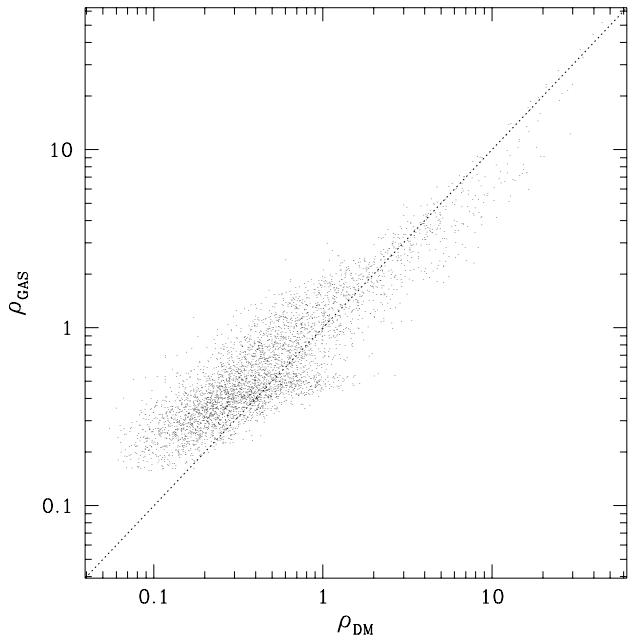
A scatter plot of the dark matter density vs the gas density
(in units of respective average densities) in the SCDM hydrodynamic
simulation at $z=3$. Because of finite gas pressure, the gas distribution does
not reach densities as low as those of the dark matter.
}
\begin{figure}
\par\centerline{%
\epsfxsize=1.0\columnwidth\epsfbox{\figname{figNH.ps}}}%
\caption{\label{figNH}\capNH}
\end{figure}
Before we move on to testing various forms of PM + initial filtering, it is
interesting to address the question of whether we need any initial smoothing at
all, i.e.\ how much the dark matter and the gas
densities differ in a hydrodynamic 
simulation. Figure \ref{figNH} shows the scatter plot of the dark matter
versus gas density for the SCDM hydrodynamic simulation at $z=3$. One can see
that the difference is significant, with the dark matter density being
a factor of 3 lower than the gas density in the lowest density regions.
Hence, a pure PM simulation, with no modifications
to mimic the dynamical effects of pressure, would
fail to reproduce the gas distribution of the low density IGM
with sufficient accuracy.

We now turn to testing the method of combining 
PM with the filtering of the initial conditions, as 
stated at the beginning of this section.
A hydrodynamic simulation is run for the SCDM model as described in
Table \ref{tabmod}. All PM simulations are performed with $64^3$ particles
on a $192^3$ mesh for the same
model. The choice of the mesh size of $192^3$ is a natural one 
given that the hydrodynamic simulation
has $64^3$ moving mesh and the softening parameter is set to $1/3$
(we have in fact tested different mesh sizes, from
$64^3$ to $256^3$, and found that the $192^3$ mesh gives, as could be expected,
much better agreement with the hydrodynamic simulation).
The pre-filtered initial conditions of the PM simulations
are chosen to be exactly the same as those in the hydrodynamic simulation.

\def\capSC{
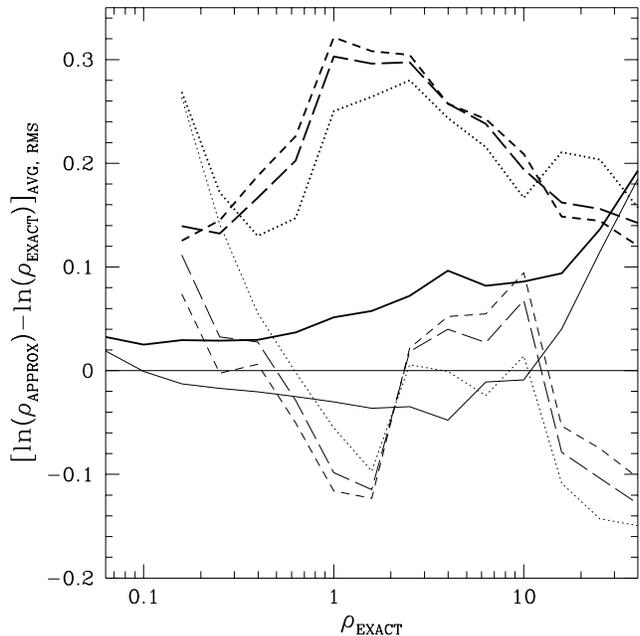
The average ({\it thin lines\/}) and rms ({\it bold lines\/}) fractional 
errors 
for the density fields in
PM + filtering simulations as compared to the 
gas density field in a full hydrodynamic 
simulation for our SCDM model (see Table 1). The different approximations
are: PM + $\exp(-k^2/k_F^2)$ smoothing
({\it dotted lines\/}), PM + $1/(1+k^2/k_F^2)$ smoothing
({\it long-dashed lines\/}), and PM + best-fit smoothing
(eq.\ \protect{\ref{bestfitfil}}; {\it short-dashed lines\/}). Also 
shown is a comparison between the dark matter density from the hydrodynamic
simulation and the density from a PM simulation without any
filtering ({\it solid lines\/}). The difference in the Green functions in the
PM and hydrodynamic simulations induces an about 5\% error.
}
\begin{figure}
\par\centerline{%
\epsfxsize=1.0\columnwidth\epsfbox{\figname{figSC.ps}}}%
\caption{\label{figSC}\capSC}
\end{figure}
Figure \ref{figSC} summarizes our findings. Before we proceed further, 
a few words are in order on our way of presenting comparisons between two
three-dimensional fields (say, density or velocity fields). 
The easiest way for such a comparison is a scatter
plot, similar to one presented in Fig.\ \ref{figNH}. However, while a scatter
plot is sufficiently illustrative, it fails to give explicit quantitative
information. 
We, therefore, use the following method of comparing 
two fields hereafter in this paper. For definiteness, suppose
we are interested in the field $Q(\bmath x)$ (which could be
density, velocity or the spectrum; in the case of spectrum, $\bmath x$ would
be replace by $\lambda$ the wavelength). 
We denote by $Q_{\rm EXACT}$ the field taken from
one of the two hydrodynamic simulations, and
by $Q_{\rm APPROX}$ the field taken from the
approximate computation under consideration.
Then we identify all spatial points in the relevant hydrodynamic simulation
which have the value of $Q_{\rm EXACT}$ within $\pm0.05{\rm dex}$ from 
some chosen value $Q_0$, and compute the following average:
\[
	\left[Q_{\rm APPROX}-Q_{\rm EXACT}\right]_{\rm AVG} \equiv
\]
\begin{equation}
	\ \ \ \ \left.\left\langle 
	Q_{\rm APPROX}(\bmath x) - Q_{\rm EXACT}(\bmath x)
	\right\rangle\right|_{Q_{\rm EXACT}(\bmath x) = Q_0}
\label{average}
\end{equation}
and the rms deviation:
\[
	\left[Q_{\rm APPROX}-Q_{\rm EXACT}\right]_{\rm RMS} \equiv
\]
\begin{equation}
	\ \ \ \ \sqrt{\left.\left\langle
	\left(Q_{\rm APPROX}(\bmath x) - Q_{\rm EXACT}(\bmath x)\right)^2
	\right\rangle\right|_{Q_{\rm EXACT}(\bmath x) = Q_0}}
\label{rms}
\end{equation}
over the ensemble of $Q_{\rm APPROX}$'s at the corresponding spatial
points in the approximate calculation. The above quantities would then
be plotted as a function of $Q_0$. 
In the case of density, we use $Q = {\rm ln}\rho$ where
$\rho$ is measured in units of the cosmic average; for the spectrum, we
use $Q = F/(1-F_{\rm EXACT})$ where $F$ is the transmission; for velocity $v$,
we use $Q = v/v_m$ where $v_m$ is defined in Figure
\ref{figVV}. In all cases, the average and rms deviations defined above
provide quantitative measures of the {\it fractional} error in the
corresponding approximate method, compared against the hydrodynamic simulation.

Given the two deviations, the average one, and the rms one,
which one is more important? The average deviation can be interpreted
as a systematic error: it measures how much the ``approximate''
density field
systematically deviates from the ``exact'' density field.
Obviously, it is desirable to reduce this error as much as
possible.
The rms deviation is more like a random error, and while it is also
desired to be as small as possible, a larger value of the random error
can perhaps be tolerated. In comparing simulations with observations, usually 
a statistical quantity is computed by averaging the results of simulations 
in some fashion. This averaging will reduce the random (rms) error,
but may not reduce the systematic (average) error. Therefore, as we
are proceeding with our tests, we will try to reduce the average error
to about 5\%, and then try to reduce the rms error as much as possible
while keeping the average error small. We again emphasize that we will
concentrate on the density range $\rho\la10$, and will ignore all
possible error induced in the high density regions.

Our ultimate object of interest is of course a comparison of the gas
distributions  
between two methods, but let us take a look at the dark matter distributions
first. There are a few interesting observations.

The agreement between the dark matter density from the hydrodynamic simulation
and the density from a PM simulation with identical initial conditions
(no pre-filtering for this PM simulation; 
shown with solid lines in Fig.\ \ref{figSC}) 
is better than 4 percent on average, and about
5\% rms, getting to 10\% at overdensities about 10. At higher overdensities the
agreement gets worse. 
What is the reason the two dark matter distributions
do not agree, in spite of the fact that the formal resolutions of two
simulations are matched? This disagreement is caused
by the difference in Green functions used to compute the gravity force in two
simulations. While the hydrodynamic simulation has the Green function  
corresponding to the Plummer softening, the PM simulations have the Green
function corresponding to our specific choice of density assignment on
the PM mesh. This difference in Green functions, which is purely
methodological, induces error of up to 10\% rms even for $\delta\sim10$
(in particular, stronger deviation at higher density is due to the fact that
the Plummer Green function is slightly softer than our PM Green function).

Finally, we turn our attention to a comparison of gas density distributions.
We first consider the simplest variant of the PM + filtering method:
we smooth the initial conditions with the $\exp(-k^2/k_F^2)$ filter (i.e. 
$\delta(k) \rightarrow \delta(k) \exp[-k^2/k_F^2]$), where
$k_F$ is given by equation (\ref{kfaskj}) with $k_J$ related to
the sound speed through (\ref{defkj}), and the evolution of the sound
speed simply taken from the hydrodynamic simulation. Recall that
this particular choice of filtering gives an excellent fit
to the exact linear baryon fluctuation on large scales (filled
circles in in Fig.\ \ref{figDB}). One might hope that the same
form of filtering + PM gives an adequate approximation even 
in the mildly nonlinear regime.

This case is shown by the dotted
lines in Fig.\ \ref{figSC}. Note that while the average error
is small for $0.5\la\rho\la10$, it gets significantly worse at 
$\rho\sim0.1$, and the rms error is as high as 20\% almost everywhere.
What causes the strong differences in low density regions?
More specifically, in such regions, why does the hydrodynamic simulation
predict gas densities substantially lower than the PM + smoothing
approximation? One possible explanation is that the choice of initial 
filtering is incorrect: the
$\exp(-k^2/k_F^2)$ filtering underestimates the amount of power at high
$k$. From the linear analysis shown 
in Fig. \ \ref{figDB}, it can be seen that this filter fails to take into
account extra power due to oscillations in the large $k$ limit. 

We therefore try two other variants of the PM + filtering method.
One is using the $1/(1+k^2/k_F^2)$ filter (shown as filled triangles in
the linear analysis of Fig.\ \ref{figDB}). 
Its results, as compared against the hydrodynamic simulation, are shown with
the long-dashed lines 
in Fig.\ \ref{figSC}. This choice of filtering gives a slower cut-off at 
high $k$ compared to the gaussian filter. The
average agreement at low densities significantly improves with this form
of filtering, at the expense of, however, increased rms error and average error
at higher densities. 

Finally, the short-dashed lines in  Fig.\ \ref{figSC} show the
results of the PM + filtering method with the following
filter function:
\begin{equation}
	f_b(t,k) = {1\over2}\left[e^{\displaystyle - k^2/k_F^2} + {1\over
	\left(1+4k^2/k_F^2\right)^{1/4}}\right].
	\label{bestfitfil}
\end{equation}
This choice of filtering gives a very good fit to the envelope of
oscillations at high $k$ in the linear fluctuations (the star symbols
in Fig.\ \ref{figDB}). One can see that the average error still reaches 10\%
within the range $\rho < 10$, and the rms error is as high as 30\% for
intermediate densities.  

We have tried quite a few other forms of filtering the initial conditions,
each giving effectively different amounts of power at high $k$, but
none of them reduces the average nor the rms error
substantially. 

We believe the fundamental flaw of the above PM + filtering procedure
is that a single uniform smoothing scale is assumed for the whole density
field. This is adequate in the linear regime where spatial fluctuations of the
temperature can be ignored in computing the filtering scale (i.e. these
fluctuations contribute to terms of higher order than those in equation
[\ref{let}]). But in the mildly nonlinear regime, one
can no longer ignore such fluctuations. In fact, places with higher 
density tends to have higher temperature \cite{HG97}, and hence higher
pressure and more smoothing. 
One then expects the lower density regions, because of their
lower thermal pressure, to be less smoothed compared to 
the higher density regions (but confined to $\rho \la 10$).
A uniform smoothing procedure would tend to overestimate the
density in the lowest density regions.
Note that the PM part of our procedure does effectively introduce 
non-uniform smoothing, but it does not do so in a way
that mimics the action of thermal pressure correctly.

We are not aware of a computationally efficient way of
performing the necessary variable smoothing on a
large mesh. Should such an algorithm 
be invented, the case for the PM + initial-filtering may be reconsidered,
but at the moment we must admit that this simple method fails to give us the
desired accuracy in reproducing 
results of the full cosmological hydrodynamic simulation, and we must
search for something better.

\section{Hydro-PM Approximation for the Cosmic Gas Distribution}

We have repeatedly emphasized in this paper that dynamically, the main
difference between dark matter and gas is that the latter is subject
to thermal pressure on top of gravity. A hydrodynamic code is designed to
compute this thermal pressure and in general, there is no other
alternative. However, in case of the
low density IGM, a very useful fact can be exploited to our
advantage: there exists a tight correlation (to better
than 10\%) between gas density and temperature (and hence pressure as
well) in the low density regime \cite{HG97}, where shock-heating is not
important. The density-temperature relation is well-described by a power-law
equation of state:
\begin{equation}
T = T_0\rho^{\gamma-1},
\label{eos}
\end{equation}
where $T_0$ is a constant of the order of $10^4 {\rm K}$, and $\gamma$ is
typically about $1.4-1.6$. 
Both $T_0$ and $\gamma$ evolve with time in a way that depends on reionization
history, but we have developed an efficient method to predict them 
with high accuracy \cite{HG97}. 

The equation of state given above immediately
provides us with the thermal pressure once the gas density is known.
The need in the hydrodynamic solver suddenly
evaporates, and the gas evolution can now be followed with a PM-type
solver, provided it is modified appropriately to include the effect
of thermal pressure. We show below how this can be done.

Let us consider the equation of motion for a cosmic gas element:
\begin{equation}
	{d{\bmath v}\over dt} + H{\bmath v} = -\nabla\phi -
	{1\over\rho}\nabla P,
	\label{eom}
\end{equation}
where ${\bmath v}$ is the gas peculiar velocity, $\phi$ is the gravitational
potential, and $P$ is the thermal pressure. If the gas is highly ionized
(so that the mean molecular weight is roughly constant, which is true
for the Lyman-alpha forest), and the temperature is
a function of density only, so that $P=P(\rho)$, equation (\ref{eom}) can be
reduced to the following equation:
\begin{equation}
	{d{\bmath v}\over dt} + H{\bmath v} = -\nabla\psi
	\label{eompsi}
\end{equation}
where
\begin{equation}
	\psi = \phi + {\cal H},
	\label{psidef}
\end{equation}
and ${\cal H}$, called {\it the specific enthalpy\/}, is
\[
	{\cal H}(\rho) = {P(\rho)\over\rho} + 
	\int_1^\rho {P(\rho^\prime)\over\rho^\prime}
	{d\rho^\prime\over\rho^\prime}.
\]
Equation (\ref{eompsi}) is identical to the equation of motion for the
collisionless dark matter except that the usual gravitational potential
$\phi$ is replaced by an effective potential $\psi$, which takes
into account both gravity and thermal pressure.
Since the gravitational
potential $\phi$ has to be computed from the density field
in a regular PM simulation anyway, it adds
only a modest computational overhead to compute
the enthalpy as well. It is extremely simple to modify
the regular PM routine to do so, and we will call this method
the ``Hydro-PM'', or HPM.

In principle, one should then follow the motion of two sets of
particles: the gas which follows the equation of motion as in (\ref{eompsi})
and the dark matter which obeys the same equation except that ${\cal H} = 0$. 
In practice, to reduce the computational cost, we
treat both sets of particles as if they all follow the same equation
of motion (equation [\ref{eompsi}], with the full $\psi$ including
both gravity and pressure). This might seem quite unjustified.
But one should bear in mind that on large scales, pressure is not
dynamically important, and so allowing pressure to also act on the dark matter
particles makes practically no difference.
The same cannot be said for small scales: the artificially imposed pressure
on dark matter causes its distribution to be less clustered than it
should be. It then becomes a question of how sensitive the
small scale pressure (which is dynamically more important than
gravity on the same scales) on the baryons is to the detailed
distribution of matter. The answer seems to be: not very much, but 
we would come back to this point in the last section.
For now, the reader can take this single component HPM method
as a plausible approximation, the merits of which can only be
weighed through detailed comparisons with 
hydrodynamic simulations. 

There is however an important technical point that we should discuss before 
going onto tests of the HPM method. In a PM (or HPM) code, the density is
assigned onto a mesh using the TSC assignment scheme, as an intermediate 
step in the computation of the potential $\phi$ (or $\psi$). As we pointed out
at the beginning of the previous section, this induces
numerical noise on small scales (high $k$). This noise is
not significant for the gravity calculation, since it is suppressed by
$k^{-2}$ power in the computation of the
gravitational potential ($\phi(k)\propto k^{-2}\delta(k)$). The computation of
the gas enthalpy, however, does not include such suppression, and the 
numerical noise could be a problem.
We therefore smooth the gas density
according to the prescription (over three mesh cells) developed at the
beginning of the previous 
section (in other words, we smooth the density field not only at the
final moment, but also at the intermediate steps of the force calculation). As
a result, the pressure force is suppressed on scales 
below about three cell sizes. It is then important that
we resolve the scale $1/k_F$ by at least three cells (assuming
the linear filtering scale $1/k_F$ gives the approximately
correct scale over which the density field is physically smoothed due to
pressure). Otherwise, the artificially reduced
pressure at scales below three cells (because of our smoothing
procedure to reduce numerical noise) could lead to unphysical 
clustering on those scales.

In the limit when the filtering scale
is very small, and is below the cell size, the pressure effect will be 
insignificant. One
then may consider running just a pure PM simulation to avoid the
additional computational expense of about 25\% because of the HPM
modification.

\def\capSH{
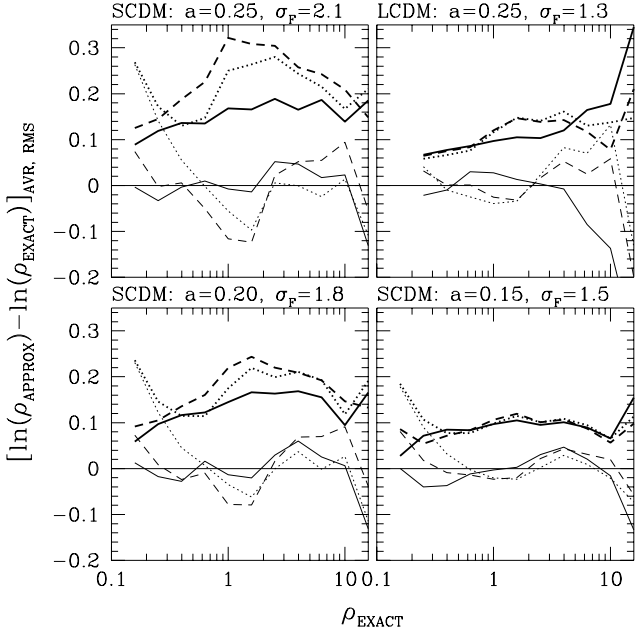
The average ({\it thin solid lines\/}) and rms ({\it bold solid lines\/}) 
fractional errors for the density fields in HPM simulations as compared to the
gas density fields in full hydrodynamic
simulations for SCDM and LCDM models, and for different stages of
evolution, as labeled for each panel. For comparison,
two variants of the PM + filtering method described in \S 3 are shown:
PM + $\exp(-k^2/k_F^2)$ smoothing ({\it thin and bold dotted lines\/} for
average and rms deviations compared with hydro) and
PM + best-fit smoothing ({\it thin and bold dashed lines\/} for average
and rms deviations compared with hydro) (see Fig.\
\protect{\ref{figSC}}). The corresponding linear rms overdensity $\sigma_F$
(eq. [{\protect{\ref{sigmaF}}}])
is also shown for each panel. 
}
\begin{figure}
\par\centerline{%
\epsfxsize=1.0\columnwidth\epsfbox{\figname{figSH.ps}}}%
\caption{\label{figSH}\capSH}
\end{figure}
Let us proceed to the comparison of the HPM approximation with full 
hydrodynamic simulations. 
We extract the equations of state as a 
function of redshift from our hydrodynamic simulations and use them
in the HPM simulation (the equations of state thus obtained agree
very well with those obtained using the method of Hui \& Gnedin 1997; 
we use for HPM the exact equations of state from the hydrodynamic
simulation so that we can focus on the error induced by
the approximate dynamics in HPM).
Figure \ref{figSH} shows the average and rms errors
for the HPM vs full hydrodynamic simulation for the SCDM model at three
different epochs and for the LCDM model at $z=3$. We also show for each panel
the corresponding 
value of $\sigma_F$, which is the rms linear overdensity for 
the $\exp(-k^2/k_F^2)$ filter:
\begin{equation}
	\sigma_F^2 = {1\over 2\pi^2}\int_0^\infty dk k^2 P_L(k,a)
	\exp(-2k^2/k_F(a)^2),
\label{sigmaF}
\end{equation}
where $P_L(k,a)$ is the linear power spectrum of a given model at a given
value of the scale factor $a$, $1/k_F(a)$ is the filtering scale at the same
moment given by equation (\ref{kfaskj}) ($\sigma_F$ grows
slower that $a$ because 
$k_F$ increases with time), and the factor of 2 in the exponential comes from
relating $\delta$ to the power spectrum by $P_L (k)\propto\delta^2(k)$. 
The quantity $\sigma_F$ therefore measures the degree of nonlinearity
of the gas distribution in the
model. At $z=3$, the SCDM model is at a more nonlinear state 
than the LCDM model.

We also show in Fig.\ \ref{figSH} two variants of the 
PM + filtering methods 
from Fig.\ \ref{figSC} for comparison.

Note that HPM gives a significantly better fit to the
gas density distribution than the PM + filtering approach. For $\delta\la10$,
the average error generally stays within 5\%, and the rms error is only weakly
dependent on density and is about 15\% for high $\sigma_F$ cases, and falling
to about 10\% for low $\sigma_F$ cases
\footnote{The LCDM model shows slightly worse
agreement at $5\la\delta\la10$. This is mostly due to the fact that we saved
fewer intermediate data while running this simulation, and as the result,
the evolution of the equation of state from this simulation is determined 
less accurately than the respective evolution from the SCDM simulation.}.
This is an important improvement over the PM + filtering method.

\def\capVV{
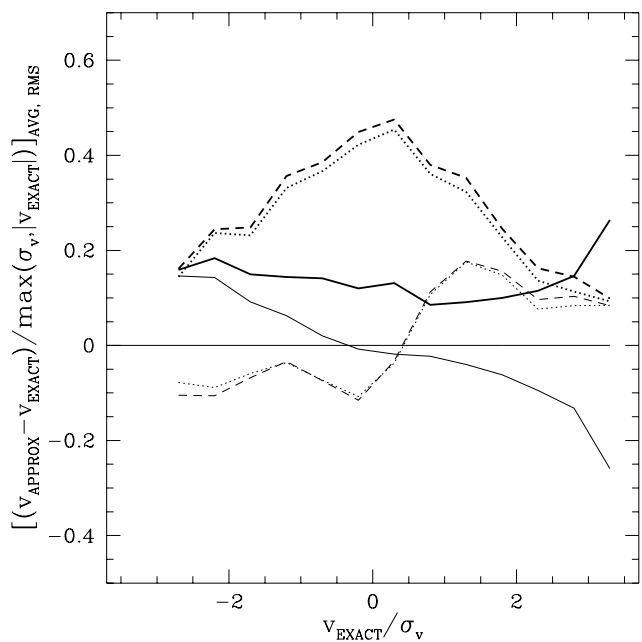
The average ({\it thin lines\/}) and rms ({\it bold lines\/}) fractional
velocity errors for: 
HPM ({\it solid lines\/}), PM + $\exp(-k^2/k_F^2)$ initial smoothing ({\it
dotted lines\/}) and PM + best-fit initial smoothing (eq.\
\protect{\ref{bestfitfil}}, {\it dashed lines\/})
as compared against the full hydrodynamic
simulation for the SCDM model at $z=3$.}

\begin{figure}
\par\centerline{%
\epsfxsize=1.0\columnwidth\epsfbox{\figname{figVV.ps}}}%
\caption{\label{figVV}\capVV}
\end{figure}
The gas density is not the only quantity that we would like to model.
For the purpose of generating absorption spectrum, it is
important that we have sufficiently accurate velocities as well.
Figure \ref{figVV} shows the comparison between one-dimensional gas
velocities (velocities projected along some fixed direction) in the HPM
approximation and in the full hydrodynamic simulation 
for our SCDM model (the solid line). 
The quantities on the y-axis in Figure \ref{figVV} are supposed
to reflect the average and rms {\it fractional} errors in the velocity.
The division by $\sigma_v$ for small $|v_{\rm EXACT}|$ is implemented
to avoid arbitrary blow up of the fractional error when the velocity
vanishes. The HPM approximation reproduces the gas velocity
again to within 15\% rms error, but the average (systematic error) has
now increased to more than 10\% for velocities in excess of two sigma.
This is an expected result, since high velocities generally correspond
to the high density regions, where the HPM approximation breaks down (because
shock-heating destroys the tight correlation between density and
temperature/pressure). 
We also show for comparison results of the PM + filtering approximations,
which cannot quite match the performance of HPM.

\def\capRRa{
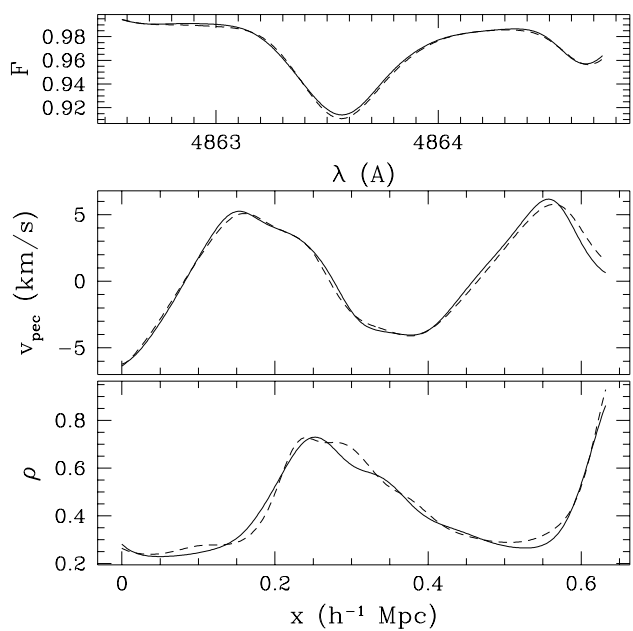
A line-of-sight comparison between a full hydrodynamic simulation ({\it solid
line\/}) and the HPM ({\it dotted line\/})
for the SCDM model at $z=3$. The bottom panel shows the density along the
line-of-sight, the middle panel shows the peculiar velocity, and the upper
panel shows the flux as a function of wavelength. This line-of-sight
goes through an underdense region.
}
\begin{figure}
\par\centerline{%
\epsfxsize=1.0\columnwidth\epsfbox{\figname{figRRa.ps}}}%
\caption{\label{figRRa}\capRRa}
\end{figure}
\def\capRRb{
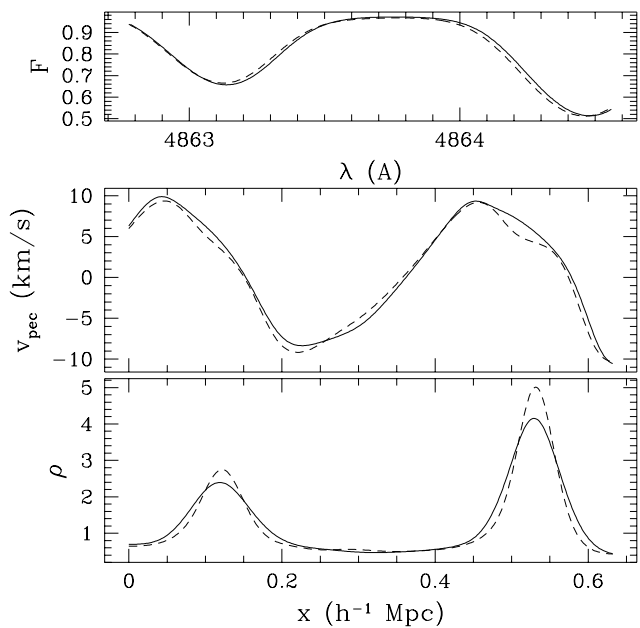
Another line-of-sight, which goes through an overdense region with $\delta<10$.
}
\begin{figure}
\par\centerline{%
\epsfxsize=1.0\columnwidth\epsfbox{\figname{figRRb.ps}}}%
\caption{\label{figRRb}\capRRb}
\end{figure}
\def\capRRc{
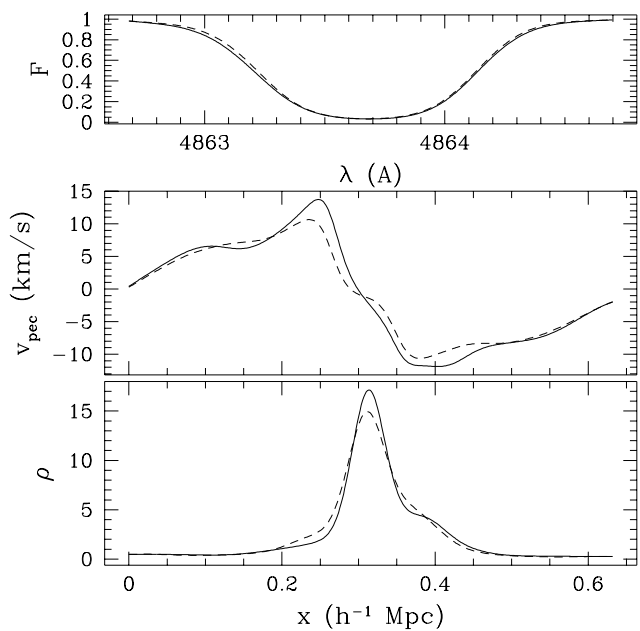
Another line-of-sight, which goes through a highly overdense region with 
$\delta>10$. The HPM approximation is expected to break down in this
regime.
}
\begin{figure}
\par\centerline{%
\epsfxsize=1.0\columnwidth\epsfbox{\figname{figRRc.ps}}}%
\caption{\label{figRRc}\capRRc}
\end{figure}
Since we plan to apply the HPM approximation to model the Lyman-alpha
forest, we must also verify that no significant systematic error is introduced
in the absorption spectra themselves. We generate spectra along randomly
oriented lines-of-sight through the hydrodynamic and the HPM simulations, and
show three examples in Figures\ \ref{figRRa}-\ref{figRRc}. 
The first line-of-sight
passes through an underdense region, the second passes through
an overdense region with overdensities $\delta\sim5$ (the HPM method
is expected to give accurate results in this case), and the third 
passes through a peak with the overdensity $\delta\sim16$. The HPM
method is expected to make a larger error in the third case, and this
can be easily observed in the corresponding bottom and middle panels, for
density and velocity
fields. However, since the transmission $F$ is related to the optical
depth ${\tau}$ by $F = e^{-\tau}$, and the optical depth is in turn
approximately proportional to density to some power, a relatively large
error in density produces only a relatively small error in $F$. 

\def\capFF{
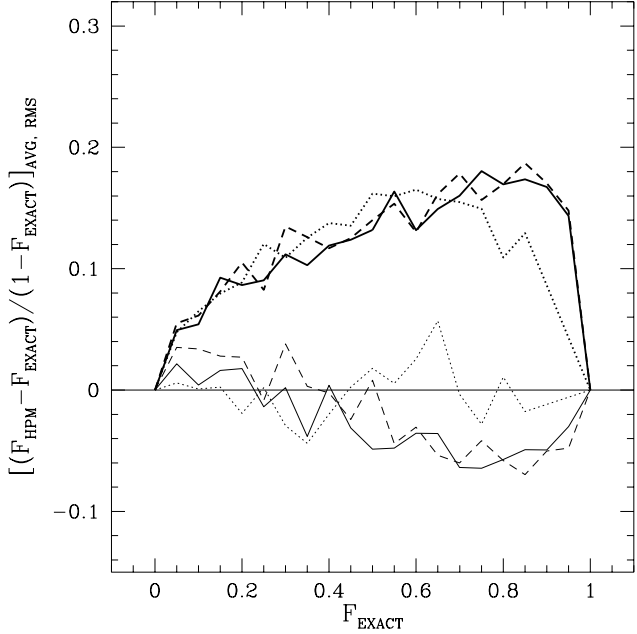
The average ({\it thin lines\/}) and rms ({\it bold lines\/}) fractional 
decrement
errors in an HPM simulation as compared to the full hydrodynamic
simulations for the SCDM model at $z=3$. The solid line
shows the HPM versus the hydrodynamic simulation for $J_{21}=0.3$, and the
dashed line shows the same comparison for $J_{21}=0.5$. Also shown is the case
when the gas temperature is decreased by a factor of 100 to reduce thermal
smoothing ({\it dotted line\/}) when generating the spectra.
}
\begin{figure}
\par\centerline{%
\epsfxsize=1.0\columnwidth\epsfbox{\figname{figFF.ps}}}%
\caption{\label{figFF}\capFF}
\end{figure}
To further quantify this, we show in Figure \ref{figFF} comparisons
between the decrements ($1-F$) in the full hydrodynamic simulation and
the HPM approximation computed from 300 random lines-of-sight. 
Since the neutral hydrogen fraction, and therefore
the decrement at a given wavelength, depends on the ionizing intensity 
$J_{21}$,
we show two different cases: $J_{21}=0.3$ (the solid line) and $J_{21}=0.5$
(the dashed line). Both values are too high for this model to reproduce
the observed column density distribution of the Lyman-alpha forest. Lower 
values of the ionizing intensity will improve the agreement, since in this
case a given value of the decrement will correspond to a lower
value of the gas density. 

One might also wonder if the above comparisons underestimate the actual error,
because of the small simulation box size:
the thermal broadening could drastically reduce discrepancies,
because the broadening width is a fair fraction of the box size
in wavelength space.
To test this possibility, we recompute the spectrum for the same lines of sight
through the HPM and the full hydrodynamic simulations with $J_{21}=0.5$, 
but with the
gas temperature reduced by a factor of 100. The corresponding comparison is
plotted in Fig.\ \ref{figFF} with the dotted line. One can see that thermal
broadening cannot explain the small errors in the transmitted flux.

Figure \ref{figFF} clearly shows the range of applicability of the HPM
approximation. While the average error stays within 10 \%, 
the rms error is smaller than about 18 \% throughout the whole
range of decrement.
A remarkable feature of the HPM approximation is that it
actually describes regions of high decrements rather well.
This is because even though the HPM method fails to give the right
density field with sufficient accuracy in high density regions,
its errors are effectively suppressed because the same regions
give rise to saturated absorption lines.

\def\capGG{
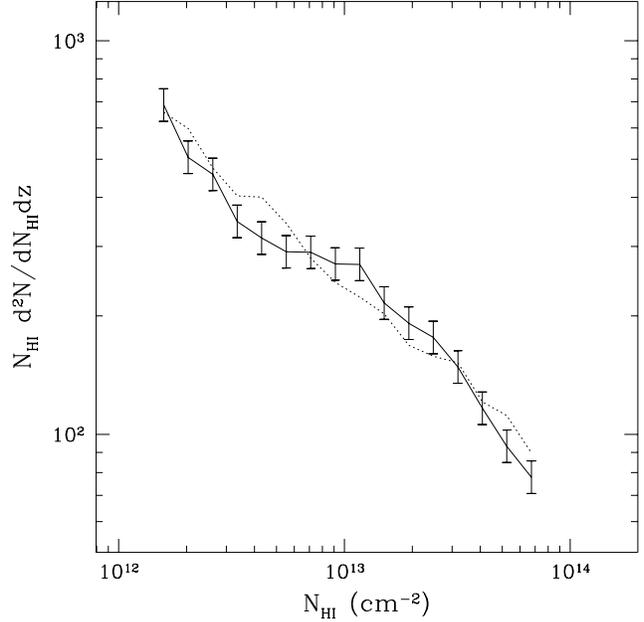
Column density distributions of the full hydrodynamic simulation ({\it solid
line\/}) and the HPM approximation ({\it dotted line\/}) for the 
SCDM model at $z=3$, computed using the Density-Peak Ansatz.
A 10\% error-bar was added for illustrative purpose only.
}
\begin{figure}
\par\centerline{%
\epsfxsize=1.0\columnwidth\epsfbox{\figname{figGG.ps}}}%
\caption{\label{figGG}\capGG}
\end{figure}
As we have emphasized before, in comparing simulations with observations,
usually  a statistical quantity is computed by averaging the results of
simulations in some fashion, which tends to reduce the random (rms) error
(in other words, the point-by-point comparisons above are a rather
stringent test).
We show one interesting example in Figure \ref{figGG}, namely the column
density distribution. We compute the column density
distributions of the full hydrodynamic simulation and the HPM
approximation 
for our SCDM model at $z=3$ using the Density-Peak Ansatz \cite{GH96,HGZ97}.
Both column density distributions are plotted in Figure \ref{figGG} with the
solid and dotted lines respectively. We also
add a 10\% error-bar to the column density distribution of the full
hydrodynamic simulation for illustrative purpose. Note that the two
distributions agree to within about 13\%, and the best-fit slopes 
differ by less than 3\%.
We thus conclude that the HPM 
approximation can be successfully used to model the Lyman-alpha forest 
when a 10-15\% accuracy is sufficient.

\section{Discussion}

We have demonstrated that the HPM method, based on a modified PM routine
to take into account the dynamical effect of pressure as well as gravity,
is an efficient and accurate alternative to hydrodynamic simulations
in predicting the density and velocity fields, as well as
absorption spectra. 

The key that makes the HPM method possible is the fact
that in the low density regime, where shock-heating is not important,
there exists a tight correlation between density and temperature/pressure.
The almost one-to-one relationship between these quantities enables
us to rewrite the equation of motion of the cosmic gas into a form
that resembles its collisionless counterpart.
The net force on the gas is then simply the gradient of an
effective potential which can be computed from the density field alone.

The power of the method is enhanced by the fact that the density-temperature
(or density-pressure) relation, which has to be input into the HPM 
computation, can be calculated for any reionization history 
in a very efficient manner without running hydrodynamic simulations 
\cite{HG97}.

We have also shown that a somewhat worse accuracy (than that of HPM) can be 
achieved by a simple
combination of a PM solver and smoothing of initial condition with an
appropriate filter. 

Both the PM + filtering method and HPM use a 
{\it single component\/} model to approximate
what is in reality a {\it two-component\/} system. 
The HPM method, in particular, treats the dark matter as if it is subject
to the same forces as the baryons, i.e.\ gravity as well as thermal pressure.
As we have explained before, this should not be a problem on large
scales, because pressure is dynamically subdominant on those scales
anyway. On small scales, we are indeed introducing an error by
allowing pressure to act on the dark matter: the dark matter distribution
would become less clustered than it should be.
One fact comes to our rescue, however: the dominant force on the baryons
on small scales is thermal pressure, not gravity, and since pressure
is determined by the baryon distribution alone, the actual
distribution of baryons on small scales should not be
sensitive to errors in the dark matter distribution.
The good agreement between results of single-component HPM and full
hydrodynamic simulations lends support to this interpretation.

We can perhaps understand this in a simpler setting. 
In Figure \ref{figDB}, the solid line shows the baryon fluctuation in the
limit $f_b=0$ (i.e.\ when no 
gravitational effect of the gas is included), while the dotted line marks the 
opposite case, $f_b=1$, (when all the matter is treated as baryonic, or in
other words, the dark matter is subjected to pressure similar to HPM). 
One might expect quite different behavior between the two cases, but in fact
they are quite similar. Both on large scales and at very small scales
(scales of oscillations), the fluctuations in both cases almost lie on
top of each other. It is at the intermediate scales, in fact close to $k_F$, 
where the two depart from each other in a perceptible way.
These scales however span a rather small range, which
is probably the reason behind the success of the single-component HPM.

Finally, a few words on the concept of maximum-likehood analysis
of the Lyman-alpha forest observations. While the HPM method is a factor
of 10-100 faster than the full hydrodynamic simulations, and only 25\%
slower than a single component PM simulation,
it still requires considerable computational expense. One can imagine
using a more efficient, but less accurate, approximation (say, truncated
Zel'dovich approximation, see Hui et al. 1997) instead of HPM.
This would introduce larger errors, but will allow us to sample a large
parameter space of cosmological models. When a smaller set of plausible models
is crudely identified with this technique, one can switch to the HPM and
further narrow the allowed parameter space to a small region; 
finally, if higher accuracy is desired, several full hydrodynamic
simulations can be run.

This work was supported in part by the 
UC Berkeley grant 1-443839-07427, and in part by the DOE and by the NASA
(NAGW-2381) at Fermilab. 
Simulations were performed on the NCSA Power Challenge Array under 
the grant AST-960015N and on the NCSA Origin2000 computer under the grant
AST-970006N.

\end{document}